\documentclass[10pt,conference]{IEEEtran}
\usepackage{graphics,graphicx,psfrag,color,float} 
\usepackage{epsfig}
\usepackage{amsmath}
\usepackage{amssymb}

\usepackage{amsmath}
\usepackage{amssymb}
\usepackage{amsfonts}
\usepackage {times}
\usepackage{braket}
\usepackage{amsfonts}

\newcommand{\beq}{\begin{equation}}
\newcommand{\enq}{\end{equation}}
\newcommand{\bel}{\begin{lemma}}
\newcommand{\enl}{\end{lemma}}
\newcommand{\bet}{\begin{theorem}}
\newcommand{\ent}{\end{theorem}}

\newcommand{\eps}{\varepsilon}
\newcommand*{\cC}{\mathcal{C}}
\newcommand*{\cA}{\mathcal{A}}

\newcommand*{\cT}{\mathcal{T}}
\newcommand*{\cX}{\mathcal{X}}
\newcommand*{\cY}{\mathcal{Y}}
\newcommand*{\cM}{\mathcal{M}}
\newcommand*{\cZ}{\mathcal{Z}}
\newcommand*{\cU}{\mathcal{U}}

\newcommand*{\cR}{\mathcal{R}}

\newcommand*{\bX}{\mathbf{X}}
\newcommand*{\bY}{\mathbf{Y}}

\newcommand*{\mximax}[1]{\Xi^\delta_{\max}(P_{#1})}
\newcommand*{\mximin}[1]{\Xi^\delta_{\min}(P_{#1})}

\newcommand*{\denc}{\mathrm{d-enc}}

\newtheorem{definition}{Definition}

\newtheorem{theorem}{Theorem}
\newtheorem{lemma}{Lemma}

\begin {document}

\title{Non-asymptotic information theoretic bound for some multi-party scenarios}
\author{
\authorblockN{Naresh~Sharma~and~Naqueeb~Ahmad~Warsi}
\authorblockA{Tata Institute of Fundamental Research\\
Homi Bhabha Road, Mumbai 400005\\
Email: \{nsharma, naqueeb\}@tifr.res.in}}

%
%
%
%
%

\maketitle

\date{\today}

\begin{abstract}
In the last few years, there has been a great interest in extending the information-theoretic
scenario for the non-asymptotic or one-shot case, i.e., where the channel is used only
once. We provide the one-shot rate region for the distributed source-coding (Slepian-Wolf)
and the multiple-access channel. Our results are based on defining a novel one-shot
typical set based on smooth entropies that yields the one-shot achievable rate regions while leveraging the results from the asymptotic analysis. Our results are asymptotically optimal, i.e., for the distributed source coding they yield the same rate region as the Slepian-Wolf in the limit of unlimited independent and identically distributed (i.i.d.) copies. Similarly for the multiple-access channel the asymptotic analysis of our approach yields the rate region which is equal to the rate region of the memoryless multiple-access channel in the limit of large number of channel uses.

\end{abstract}

\maketitle

\section{Introduction}

Over the last several decades, information theory has been used to analyze the performance of various information processing
tasks such as data compression, information transmission across a noisy communication channel etc. The analysis
has traditionally been asymptotic wherein a certain task is repeated unlimited number of times. Such an approach
has yielded rich dividends providing fundamental limits to the performance and gave operational meaning to quantities
such as entropy and mutual information \cite{shannon1948, covertom}.

But the assumption of repeating an information processing task is hardly realistic. There has been a great interest
over the last few years to analyze a task done only \emph{once}. Typically, the analysis involves finding lower and upper
bounds on the resources needed to perform such a task. This has been referred to as ``one-shot" or ``single-shot" in
the literature.

The lower and upper bounds are expected to be asymptotically tight, i.e., they both yield the same average number of resources
needed per task when the number of times a task is performed is unbounded and this quantity is also equal to the asymptotic
analysis that has been done traditionally. It is interesting to note that while the asymptotic analysis would start with the asymptotic
equipartition property (AEP), the one-shot analysis could, at the very end, be augmented with AEP to yield the same answer.
Furthermore, one-shot analysis has been applied to the cases where i.i.d. (independent and identically distributed) and the
memoryless assumption is not valid. For example, in a data compression task, we may not have i.i.d. copies of the
random variable modeling the source output or in a communication across a noisy channel, the channel may not be memoryless.

The one-shot bounds have been given more recently by Polyanskiy et al. in Ref. \cite{Polyanskiy-Poor-Verdu-2012} and Renner et al. in Ref. \cite{renner-wolf-2004,renner-wolf-2005}. In Ref. \cite{renner-wolf-2004,renner-wolf-2005} Renner et al. introduced the elegant notion of smooth R\'{e}nyi entropies (defined later) where
they also gave one-shot bounds for data compression and randomness extraction (see also Ref. \cite{cachin-isit-1997}).
Another important application of smooth R\'{e}nyi entropy measure has been the one-shot bounds for the channel
capacity given by Renner, Wolf and Wullschleger \cite{renner-class-cap-2006}. Channel coding bounds are also derived in
Ref. \cite{renner-isit-2009} using a quantity called the smooth $0$-divergence, which is a generalization of
R\'{e}nyi's divergence of order $0$. To the best of our knowledge for the point to point systems Polyanskiy et al's. approach has come out better against the smooth entropy based approach in all the numerical studies done thus far.

The smooth R\'{e}nyi entropies were extended to the quantum case by Renner and K\"{o}nig \cite{renner-konig-2005}.
There has been a considerable work on the one-shot bounds for the quantum case under various scenarios (see for
example Refs. \cite{datta-renner-2009, konig-op-mean-2009,dupuis-broadcast-2010,berta-reverse-shannon-2011,
datta-fqsw-2011,renes-renner-2011} and references therein).

In Ref. \cite{schoe-isit-2007},  Schoenmaker \emph{et al} introduce the notion of
smooth R\'{e}nyi entropy for the case of  stationary ergodic information sources, thereby generalizing
previous work which concentrated mainly on i.i.d. information sources.
In Ref. \cite{holenstein-renner-2011}, Holenstein and Renner give explicit and tight bounds on
the smooth entropies of $n$-fold product distributions in terms of the Shannon entropy of a single distribution. Renes and Renner
consider the problem of classical data compression in the presence of quantum side information in Ref. \cite{renes-renner-2010}. 

It is worth noting that for the classical case, our results (announced in Jan 2012) were the first one-shot multi-party results,
i.e., where there are multiple parties at the sender and/or receiver. After our results were were announced on the arxiv, Tan and Kosut also
gave non-asymptotic bounds for several multi-party problems
following the dispersion based approach \cite{Vincent-Tan-2012}. In this paper
we concentrate on the distributed encoding (data compression) protocol of correlated information sources
and for finding the one-shot transmission rate. We then give one-shot achievable rate pair for the multiple-access channel.
In a very remarkable and fundamental paper, Slepian and Wolf showed that there is no loss in the compression 
efficiency for the distributed encoding as compared to the collaborative encoding \cite{slepian-wolf-1973}. In Ref. \cite{Han}, Han proves a more general result wherein he gives the asymptotically optimal rate region for distributed source coding in the non i.i.d. regime.

The fundamental contributions by Renner and Wolf
\cite{renner-wolf-2004,renner-wolf-2005} have not been extended thus far to the multi-party
scenario. The upper bound that Renner and Wolf derive for their data compression comes in the form of two
steps. One is the encoding (that can be accomplished
by two-universal hash functions) and second is an elegant step of bootstrapping the first step
by encoding a random variable that is in a ball (appropriately defined) around the random variable that
we want to compress. Such an approach distributes the error in the two steps in a flexible way and yields an upper bound
that is asymptotically tight.
To carry over this approach to the Slepian-Wolf protocol would involve proving the existence of more than one random
variables that satisfy certain conditions and the authors know of no such proof in the literature.

In this paper, we define a one-shot typical set. We further use this set to give one-shot achievable rates for some multi-party scenarios. Interestingly, our bounds involves smooth R\'{e}nyi entropy of the order $-\infty$. R\'{e}nyi entropies are
typically defined of order $\alpha \geq 0$ and appropriate limits have to be taken for $\alpha = 1$ that corresponds
to the Shannon entropy \cite{renyi-1960}. In the definition of smooth R\'{e}nyi entropy, we let $\alpha \in [-\infty,\infty]$.

\section{Definition of various entropies}

We shall assume that all random variables in this paper are discrete and take values over a finite set.
Let $X$ be a random variable taking values over
alphabet $\cX$ with probability mass function (PMF) $P_X(x)$, $x \in \cX$. Then
the Shannon entropy \cite{shannon1948} $H(X)$ of $X$ is defined by 
\beq
H(X) := - \sum_{x \in \cX}P_X(x)\log [P_X(x)],
\enq
where we shall assume that the log is to the base $2$ throughout this paper.
The R\'{e}nyi entropy \cite{renyi-1960} of $X$ is defined as 
\beq
H_{\alpha}(X) := \frac{1}{1-\alpha}\log \left[ \sum_{x\in \cX} P^{\alpha}_X(x) \right],
\enq
where $\alpha \in [0,\infty]$ and appropriate limits are taken for $\alpha = 1$. In particular,
for $\alpha = 0$, the R\'{e}nyi entropy is given by 
\beq
H_{0}(X) := \log|\{x \in \cX: P_{X}(x) > 0 \}|
\enq
and for $\alpha = \infty$, the R\'{e}nyi entropy is given by
\beq
H_{\infty}(X) := -\log \left[ \max_{x\in \cX} P_X(x) \right].
\enq
Let $X,Y$ be two random variables with joint distribution $P_{XY}$,
the conditional R\'{e}nyi entropy of order $\alpha$ is defined as
\beq
H_{\alpha}(X|Y) := \frac{1}{1-\alpha}\log \left[ \max_{y \in \cY} \sum_{x\in \cX} P^{\alpha}_{X|Y=y}(x) \right].
\enq
The $\eps$-smooth conditional R\'{e}nyi entropy of order $\alpha$ \cite{renner-wolf-2004} is defined for $\eps\geq 0$ as
\begin{align}
\label{halp-cond}
H_{\alpha}^{\eps}(X|Y) &:=  \frac{1}{1-\alpha}\log \bigg[  \inf_{\bar{X} \bar{Y} : \Pr[\bar{X} \bar{Y} \neq XY] \leq \eps} \max_{y \in \cY} \nonumber\\
& \hspace{25mm} \sum_{x\in \cX} P^{\alpha}_{\bar{X}|\bar{Y}=y}(x) \bigg].
\end{align}

Although the R\'{e}nyi entropy is typically defined for $\alpha \in [0,\infty]$, we let $\alpha \in [-\infty,\infty]$
and, in particular, consider the R\'{e}nyi entropy of order $\alpha = -\infty$ given by
\begin{align}
H_{-\infty}(X|Y) &:= -\log \bigg[ \min_{y \in \cY} ~ \min_{x : P_{X|Y=y}(x) > 0}\nonumber \\
& \hspace{20mm} P_{X|Y=y}(x) \bigg] 
\end{align}
and its smooth version for $\eps \geq 0$ as
\begin{align}
H_{-\infty}^\eps(X|Y) &:= -\log \bigg[ \sup_{\bar{X} \bar{Y} : \Pr[\bar{X} \bar{Y} \neq XY] \leq \eps} \min_{y \in \cY}\nonumber\\
& \hspace{15mm} \min_{x : P_{\bar{X}|\bar{Y}=y}(x) > 0} P_{\bar{X}|\bar{Y}=y}(x) \bigg].
\end{align}

We shall use $H_\alpha^\eps(X)$ interchangeably with $H_\alpha^\eps(P_X)$.

\section{Distributed encoding of correlated sources}

We describe the task of distributed encoding of correlated sources in this section.
Let $(X,Y) \sim P_{XY}$. Assume that the random variable $X$ is available with
Alice at a location $A$ and the random variable $Y$ is available with Bob at a
separate location $B$. Both Alice and Bob want to get across the pair $(X,Y)$ to Charlie
at location $C$ without collaborating with each other within some prescribed error. We state this formally as follows.

\begin{definition}
For the error $\eps$, $0 \leq \eps \leq 1$,
let $\Lambda^P_{\eps}(\cX \times \cY \rightarrow \cC_\cX \times \cC_\cY)$ 
denote the set $(e_\cX,e_\cY,U)$ where
$U$ is a random variable with range $\cU$, $e_\cX : \cX \times \cU \rightarrow \cC_\cX$ and $e_\cY : \cY \times \cU \rightarrow \cC_\cY$
such that there exists a decoding function $g : \cC_\cX \times \cC_\cY \times \cU \rightarrow \cX \times \cY$ with
\begin{align}
\label{sw-err}
\Pr\left\{ (\hat{X}(U), \hat{Y}(U)) \neq (X,Y) \right\} & \leq \eps,
\end{align}
where
\beq
(\hat{X}(U),\hat{Y}(U)) := g \left[ e_\cX(X,U), e_\cY(Y,U), U \right].
\enq
Occasionally, we shall also use the following
\begin{align}
g_\cX \left[ e_\cX(X,U), e_\cY(Y,U), U \right] & := \hat{X}(U), \\
g_\cY \left[ e_\cX(X,U), e_\cY(Y,U), U \right] & := \hat{Y}(U).
\end{align}

The definitions for the encoding lengths for a chosen $(e_\cX,e_\cY,U) \in \Lambda^P_{\eps}$
are given by
\begin{align}
\label{xenc}
\ell_{\denc}^{\eps}(X) & := \log |\cC_\cX|, \\
\ell_{\denc}^{\eps}(Y) & := \log |\cC_\cY|,
\end{align}
where we have written $\Lambda^P_{\eps}$ for
$\Lambda^P_{\eps}(\cX \times \cY \rightarrow \cC_\cX \times \cC_\cY)$
and we would implicitly assume that $\cC_\cX$ and $\cC_\cY$ are determined
from the same $(e_\cX,e_\cY,U)$. We also assume that $U$ is independent of
$X$ and $Y$.
\end{definition}

\subsection{Results}

In this section, we summarize the results and the proofs are given in the next section.
We first need the following definition.
\begin{definition}
For $(X,Y) \sim P_{XY}$, we define the following sets
\begin{align}
\label{const. X}
\cA_{\delta}(P_X) & := \{ x : \mximin{X} \leq   -\log P_X(x) \leq \nonumber \\
& \hspace{8mm}  \mximax{X} \},
\end{align}
where
\begin{align}
\mximin{X} & := H(X) - |H(X) - H_{\infty}^{\delta}(X)| \nonumber\\
& \hspace{5mm} - \delta \log|\cX|, \\
\mximax{X} & := H(X) + | H(X) - H_{-\infty}^{\delta}(X) | \nonumber \\
& \hspace{5mm} + \delta \log|\cX|,
\end{align}
\begin{align}
\cA_{\delta}(P_Y) & := \{y : \mximin{Y} \leq -\log P_Y(y)  \leq \nonumber \\
 & \hspace{8mm} \mximax{Y}\}, \\
\label{const. XY}
\cA_{\delta}(P_{XY}) & := \{(x,y) : \mximin{XY}  \leq -\log P_{XY}(x,y)  \nonumber \\
& \hspace{7mm} \leq  \mximax{XY},  x\in \cA_{\delta}(P_X), y \in \nonumber \\ 
& \hspace{8mm}\cA_{\delta}(P_Y) \}.
\end{align}
\end{definition}

We now give the one-shot bounds for the distributed encoding for correlated sources in the following theorem.
\begin{theorem}
\label{sw}
Let $(X,Y) \sim P_{XY}$ and $\eps$, $\eps_i \in \mathbb{R}^{+}$, $i=0,1,2,3$ with $\sum_{i=0}^3 \eps_i \leq \eps$. \\
\noindent {\bf Lower bounds}:
For any distributed encoding protocol, the following lower bounds shall hold
\begin{align}
\label{lbxsw}
\ell_{\denc}^{\eps}(X) & \geq H_0^{\eps}(X|Y), \\
\label{lbysw}
\ell_{\denc}^{\eps}(Y) & \geq H_0^{\eps}(Y|X), \\
\label{lbxysw}
\ell_{\denc}^{\eps}(X) + \ell_{\denc}^\eps(Y) & \geq H_0^{\eps}(X,Y).
\end{align}
{\bf Achievability}: There exists a distributed encoding protocol for a rate pair \linebreak
$(\ell_{\denc}^{\eps}(X),\ell_{\denc}^{\eps}(Y))$
satisfying the following conditions
\begin{align}
\label{axsw}
\ell_{\denc}^{\eps}(X) & \geq \mximax{XY} - \mximin{Y} \nonumber \\
& \hspace{5mm} - \log\eps_1, \\
\label{aysw}
\ell_{\denc}^{\eps}(Y) & \geq \mximax{XY} - \mximin{X} \nonumber \\
& \hspace{5mm} - \log\eps_2, \\
\label{axysw}
\ell_{\denc}^{\eps}(X) + \ell_{\denc}^\eps(Y) & \geq \mximax{XY} - \log\eps_3,
\end{align}
where $\delta$ is chosen such that
\beq
\Pr \left\{ (X,Y) \notin \cA_\delta(P_{XY}) \right\} \leq \eps_0.
\enq
\end{theorem}

Next we show that one-shot rate region in Theorem \ref{sw} is asymptotically
optimal, i.e., both the lower bounds and the achievable rates (normalized appropriately) are the same when the number of i.i.d. copies of
$(X,Y)$ pairs grows unbounded and the errors become arbitrarily small, and that this rate region in the distributed
encoding case comes out to be the same as that of Slepian-Wolf \cite{slepian-wolf-1973}. Let
\beq
\label{vec-def}
(\bX_1^n,\bY_1^n) := [(X_1,Y_1), (X_2,Y_2),...,(X_n,Y_n)],
\enq
where $(X_1,Y_1),...(X_n,Y_n)$ are $n$ i.i.d. pairs of random variables distributed according to $P_{XY}$.

Similarly, we can get the asymptotic limit for Theorem \ref{sw} which yields the rate-region of Slepian-Wolf \cite{slepian-wolf-1973}. Before we get into that, 
we first give the definition of rate pair which are achievable asymptotically and this definition is motivated from Ref. \cite{Han}.
\begin{definition}
\label{rate}
Rate pair $(\cR_1, \cR_2)$ is achievable $\iff$ There exists a
\beq
(e_{\cX^n},e_{\cY^n},U) \in \Lambda^{P_{\bX_1^n \bY_1^n}}_{\eps}
(\cX^n \times \cY^n \rightarrow \cC_{\cX^n} \times \cC_{\cY^n})
\enq
and a decoding function $g : \cC_{\cX^n} \times \cC_{\cY^n} \times \cU \rightarrow
\cX^n \times \cY^n$ such that
\begin{align*}
\Pr\left\{g \left[ e_{\cX^n}(X_1^n,U), e_{\cY^n}(Y_1^n,U), U \right]\neq (X_1^n,Y_1^n)\right\}
& \leq \eps, \\
\lim_{\eps \to 0}\limsup_{n \to \infty}\frac{ \ell_{\denc}^{\eps}(\bX_1^n)}{n} & \leq \cR_1 \\
\lim_{\eps \to 0}\limsup_{n \to \infty}\frac{ \ell_{\denc}^{\eps}(\bY_1^n)}{n} & \leq \cR_2.
\end{align*}
\end{definition}

\begin{lemma}
\label{asymp-sw}
The normalized asymptotic limits for the bounds in Theorem \ref{sw}, are given by
\begin{align}
\cR_1 \geq H(X_1|Y_1), \\
\cR_2 \geq H(Y_1|X_1), \\
\cR_1 + \cR_2 \geq H(X_1,Y_1).
\end{align}
\end{lemma}

\subsection{Proofs of the results}

Some of the proofs rely upon the two-universal hash functions and we give their definition first (see also Ref. \cite{bennett-1995}
and references therein).

\begin{definition}
A random function $f : \cX \times \cU \to \cC$ takes an input $x \in \cX$ and generates a uniform random variable $U$ taking values
over alphabet $\cU$ and outputs $c \in \cC$. $f$ is called a two-universal
hash function if for any $x \neq \acute{x}$, $x, \acute{x} \in \cX$, we have
\beq
\Pr \left\{ f(x,U) = f(\acute{x},U) \right \} \leq \frac{1}{|\cC|}.
\enq
\end{definition}

\noindent We prove the following lemma that shall be needed for the probability of error analysis later.
\begin{lemma}
\label{a-prop}
Let $(X,Y) \sim P_{XY}$. Define
\beq
\cA_{\delta}(P_{X|Y=y}) := \{x : (x,y) \in \cA_{\delta}(P_{XY})\}.
\enq
For any $\delta \geq 0$, $y \in \cA_{\delta}(P_Y)$, we have
\beq
\label{A_x_given_y}
|\cA_{\delta}(P_{X|Y=y})|\leq 2^{\mximax{XY} - \mximin{Y}}.
\enq
\end{lemma}
\begin{proof}
For any $(x,y)\in \cA_{\delta}(X,Y)$, we have
\beq
\label{temp1}
P_{X|Y=y}(x) = \frac{P_{XY}(x,y)}{P_Y(y)} \geq 2^{-\mximax{XY} + \mximin{Y}}.
\enq
The lemma now follows straightforwardly from
\begin{align}
1 &\geq \sum_{x\in \cA_{\delta}(X|Y=y)}P_{X|Y=y}(x)\\
& \geq |\cA_{\delta}(X|Y=y)| 2^{-\mximax{XY} + \mximin{Y}},
\end{align}
where we have used \eqref{temp1}.
\end{proof}

The following properties of $H_{-\infty}^\eps$ are needed later in the paper.
\begin{lemma}
\label{minf-prop1}
The following holds for any $\eps > 0$
\begin{align}
H_{-\infty}^\eps(X) & \geq H_0^\eps(X).
\end{align}
\end{lemma}
\begin{proof}
For all $\delta > 0$, there exists a random variable $\bar{X}$ with $\Pr\{X \neq \bar{X}\} \leq \eps$ such that
\beq
H_{-\infty}^\eps(X) \geq H_{-\infty}(\bar{X}) - \delta.
\enq
Now the following inequalities hold
\begin{align}
H_{-\infty}^\eps(X) & \geq H_{-\infty}(\bar{X}) - \delta \\
& \geq H_0(\bar{X}) - \delta \\
& \geq H_0^{\eps}(X) - \delta.
\end{align}
Since this is true for any $\delta > 0$, the result follows.
\end{proof}

\begin{lemma}
\label{minf-prop2}
If $X_1^n = [X_1,...,X_n]$ be $n$ i.i.d. random variables distributed according to $P_X$, then
\beq
\lim_{\eps \to 0} \lim_{n \to \infty} \frac{H_{-\infty}^{\eps}(X_1^n)}{n} = H(X_1).
\enq
\end{lemma}
\noindent
\begin{proof} The typical set, $\cT_\eps^{(n)}$ $\subseteq \cX^n$, $\eps \geq 0$ (see Ref. \cite{covertom} for details) is defined as
\beq
\cT_\eps^{(n)} = \left\{ x^n : \left|-\frac{1}{n}\log P(x_1^n) - H(X_1) \right| \leq \eps \right\}.
\enq
We now define a random variable $Z$ as
\beq
Z =
\begin{cases}
X_1^n & \mbox{if } X_1^n \in \cT_\eps^{(n)} \\
W         & \mbox{if } X_1^n \notin \cT_\eps^{(n)}
\end{cases}
\enq
where $W$ is uniformly chosen at random from the set $\cT_\eps^{(n)}$. We now have 
\begin{align}
\Pr \left\{ Z \neq X_1^n \right\} &=  \Pr \{ X_1^n \notin \cT_\eps^{(n)} \} \\
\label{temp2}
                                                    & \leq \eps,
\end{align}
where we have assumed that $n$ is large enough so that \eqref{temp2} is satisfied.
To get a bound on $\min_{z} \Pr\{Z = z\}$, we note that for any $x_1^{n} \in \cT_\eps^{(n)}$, $\Pr\{X_1^n = x_1^{n}\}$
$\geq 2^{-n(H(X_1)+\eps)}$.
Now using the definition of $H_{-\infty}^\eps(X)$ and $\Pr\{ Z \neq X_1^n \} \leq \eps$, we have
$H_{-\infty}^\eps(X_1^n) \leq H_{-\infty}(Z)$, and hence,
\beq
\frac{H_{-\infty}^{\eps}(X_1^n)}{n} \leq  \frac{H_{-\infty}(Z)}{n}\leq H(X_1) + \eps,
\enq
or
\beq
\label{eq:upper}
\lim_{\eps \to 0} \lim_{n \to \infty} \frac{H_{-\infty}(X_1^n)}{n} \leq H(X_1).
\enq
We now use Lemma \ref{minf-prop1} to get $H_{-\infty}^{\eps}(X_1^n) \geq H_0^{\eps}(X_1^n)$ and hence,
\beq
\label{eq:lower}
\lim_{\eps \to 0} \lim_{n \to \infty} \frac{H_{-\infty}^{\eps}(X_1^n)}{n} \geq 
\lim_{\eps \to 0} \lim_{n \to \infty} \frac{H_0^{\eps}(X_1^n)}{n} = H(X_1),
\enq
where the equality in \eqref{eq:lower} follows from Ref. \cite{renner-wolf-2005}.
Thus, from \eqref{eq:upper} and \eqref{eq:lower}, the lemma is proved.
\end{proof}

%

\subsection{Proof of the lower bound in Theorem \ref{sw}}

Let $(e_\cX,e_\cY,R) \in \Lambda^P_{\eps}(\cX \times \cY \rightarrow
\cC_\cX \times \cC_\cY)$. Then there must exist a realization $R=r$ such that the error bound in \eqref{sw-err} is met.
We first prove \eqref{lbxsw}.
\begin{align*}
\ell_\denc^{\eps}(X) & \overset{a}{=} H_0[e_\cX(X,r)] \\
& \overset{b}{\geq} H_0[e_\cX(X,r) | Y] \\
& \overset{c}{=} H_0[e_\cX(X,r), e_\cY(Y,r) | Y] \\
& \overset{d}{\geq} H_0 \left\{ g_\cX \left[ e_\cX(X,r), e_\cY(Y,r),r \right] | Y \right\} \\
& = H_0 [ \hat{X}(r) | Y ] \\
& \overset{e}{\geq} H_0^{\eps}(X|Y),
\end{align*}
where $a$ follows from the definition of $\ell_\denc^{\eps}(X)$, $b$ follows since conditioning reduces $H_0$, $c$ is an identity
that is easily derived from the definition of $H_0$, $d$ follows since taking a function reduces $H_0$, and $e$ follows since
$\Pr\{\hat{X}(r) \neq X \} \leq \eps$ and from \eqref{halp-cond}.

The proof for \eqref{lbysw} is similar and is omitted. We now prove \eqref{lbxysw}.
\begin{align*}
\ell_\denc^{\eps}(X) + \ell_\denc^{\eps}(Y) & = H_0[e_\cX(X,r)] + H_0[e_\cY(Y,r)] \\
& \overset{a}{\geq} H_0[e_\cX(X,r), e_\cY(Y,r)] \\
& \geq H_0 \left\{ g \left[ e_\cX(X,r), e_\cY(Y,r),r \right] \right\} \\
& = H_0 [ \hat{X}(r), \hat{Y}(r) ] \\
& \geq H_0^{\eps}(X,Y),
\end{align*}
where $a$ follows from the sub-additivity of the R\'{e}nyi entropy.

\subsection{Proof of achievability in Theorem \ref{sw}}

We show the existence of a protocol for distributed encoding (again based on two-universal hash functions) and is given by
the following steps.

\begin{enumerate}
\item Let $e_\cX : \cX \times \cU \to \cC_\cX$, $e_\cY : \cY \times \cU \to \cC_\cY$ be two-universal hash functions and let
${\cC_\cX} = \{1,2,\cdots,2^{\ell_\cX}\}$ and ${\cC_\cY} = \{1,2,\cdots,2^{\ell_\cY}\}$.
Alice takes $x \in \cX$, generates $U$ (known both to Bob and Charlie),
and sends $i = e_\cX(x,U)$ to Charlie.
Similarly, Bob takes $y \in \cY$ and sends $j = e_\cY(y,U)$ to Charlie.

\item Charlie passes the received indices $(i,j)$ to the decoder $g: \cC_\cX \times \cC_\cY \times \cU \to \cX \times \cY$
that outputs $(\hat{x},\hat{y}) = (x,y)$ if there is only one pair $(x,y)$ such that $e_\cX(x,U) = i$, $e_\cY(y,U) = j$
and $(x,y) \in \cA_{\delta}(P_{XY})$. In all other cases, it declares an error.
\end{enumerate}

Let $\eps_0, \eps_1, \eps_2, \eps_3 \in \mathbb{R}^+$ and $\sum_{k=0}^3 \eps_k \leq \eps$.
The probability of error of the above protocol is computed by defining the following events.
\begin{align}
E_0 & := \{(X,Y) \notin \cA_{\delta}(P_{XY}) \},\\ 
E_1 & := \{\exists ~ \acute{x} \neq X : e_\cX(\acute{x},U) = e_\cX(X,U), (\acute{x},Y) \nonumber \\
& \hspace{6mm} \in \cA_{\delta}(P_{XY}) \}. \\
E_2 & := \{\exists ~ \acute{y} \neq Y : e_\cY(\acute{y},U) = e_\cY(Y,U), (X,\acute{y}) \nonumber\\
&\hspace{6mm}  \in \cA_{\delta}(P_{XY}) \}. \\
E_{12} & := \{\exists (\acute{x},\acute{y}) : (\acute{x},\acute{y}) \neq (X,Y), e_\cX(\acute{x},U) =\nonumber \\ 
& \hspace{7mm} e_\cX(X,U), e_\cY(\acute{y},U) = e_\cY(Y,U), (\acute{x},\acute{y})\nonumber \\
& \hspace{6mm} \in \cA_{\delta}(P_{XY})\}.
\end{align}
The probability of error is given by the probability of the union of these events and we upper bound that by the
union bound as
\begin{align}
P_e & = \Pr \left\{ E_0 \cup E_1 \cup E_2 \cup E_{12} \right\} \\
         & \leq \Pr\{ E_0 \} + \Pr\{ E_1 \} + \Pr\{ E_2 \} + \Pr\{ E_{12} \}.
\end{align}
For further details on the calculation of $P_e$ and the proofs of \eqref{axsw}, \eqref{aysw} and \eqref{axysw} see
Ref. \cite{sharma-warsi-2011}.\\

\subsection{Proof of Lemma \ref{asymp-sw}}

The proof of Lemma \ref{asymp-sw} follow straightforwardly if we could invoke Lemma \ref{minf-prop2},
which we can if we can show that $\delta$ is not zero for large enough $n$. It is here that we shall need the following
remarkable result by Holenstein and Renner \cite{holenstein-renner-2011} giving explicit bounds.
\begin{theorem}(Holenstein and Renner \cite{holenstein-renner-2011})
Let $P_{\bX^n_1 \bY^n_1} := P_{X_1 Y_1} \cdots P_{X_n Y_n}$ be a probability distribution over $\cX^n \times \cY^n$.
Then, for any $\delta \in [0,\log |\cX| ]$ and ${\mathbf x}, {\mathbf y}$ chosen according to $P_{\bX^n_1 \bY^n_1}$
\beq
\underset{{\mathbf x}, {\mathbf y}}{\Pr} \left[ -\log ( P_{\bX^n_1 | \bY^n_1} ({\mathbf x}, {\mathbf y}) ) \geq
H(\bX^n_1 | \bY^n_1) + n \delta \right] \leq \eps \nonumber
\enq
and, similarly
\beq
\underset{{\mathbf x}, {\mathbf y}}{\Pr} \left[ -\log ( P_{\bX^n_1 | \bY^n_1} ({\mathbf x}, {\mathbf y}) ) \leq
H(\bX^n_1 | \bY^n_1) - n \delta \right] \leq \eps \nonumber
\enq
where $\eps = 2^{- \frac{n \delta^2}{2 \log^2(|\cX|+3)}}$.
\end{theorem}

Let us now assume that we want to ensure that for $\delta = \eps_0$,
\beq
\label{dummy1}
\Pr\{ \bX_1^n \notin A_\delta(P_{\bX_1^n}) \} \leq \eps_0.
\enq
Recall that $X_1,X_2,...$ are i.i.d. and distributed according to $P_X$.
Then it is sufficient to ensure that
\begin{align*}
\underset{{\mathbf x}}{\Pr} \left[ -\log ( P_{\bX^n_1} ({\mathbf x}) ) \geq nH(X_1) + n \delta \log|\cX| \right] & \leq \frac{\eps_0}{2}, \\
\underset{{\mathbf x}}{\Pr} \left[ -\log ( P_{\bX^n_1} ({\mathbf x}) ) \leq nH(X_1) - n \delta \log|\cX| \right] & \leq \frac{\eps_0}{2},
\end{align*}
where we have taken out the term inside $|\cdot|$ to get upper bounds to the probabilities. Now invoking the result by
Holenstein and Renner \cite{holenstein-renner-2011}, if we choose $\delta = \eps_0$, then for all $n \geq n_0(\eps_0)$, where
\beq
n_0(\eps_0) := \sqrt{ - \frac{2 \log(\eps_0/2) \, \log^2(|\cX|+3)}{ \log^2(|\cX|) \, \eps_0^2} },
\enq
the error bound in \eqref{dummy1} is met. Note that we could choose $\delta = 0$ for $n < n_0(\eps_0)$. Hence, both the sequences
$\{ H_{-\infty}^\delta(X_1^n)/n \}$ and $\{ H_{-\infty}^{\eps_0}(X_1^n)/n \}$ are equal $\forall$ $n \geq n_0(\eps_0)$. Hence,
\begin{align*}
\lim_{\eps \to 0} \lim_{n \to \infty} \frac{H_{-\infty}^{\delta}(X_1^n)}{n} &= \lim_{\eps \to 0} \lim_{n \to \infty} \frac{H_{-\infty}^{\eps_0}(X_1^n)}{n}\\
& \hspace{1mm}= \lim_{\eps_0 \to 0} \lim_{n \to \infty} \frac{H_{-\infty}^{\eps_0}(X_1^n)}{n}.
\end{align*}
Now Lemma \ref{minf-prop2} can be invoked to yield the desired answer.
It is straightforward to extend the above argument for the i.i.d. sequence of a pair of
random variables $(X,Y)$.

Since, the bounds derived in the achievability part of Theorem \ref{sw} are true for any $n$, therefore using Definition \ref{rate} and the above discussion we get the following bounds in the asymptotic regime for any achievable rate pair $(\cR_1, \cR_2)$
\begin{align*}
\cR_1\overset{a} \geq \lim_{\eps \to 0}\limsup_{n \to \infty} \frac{ \ell_{\denc}^{\eps}(\bX_1^n)}{n}  \geq H(X_1|Y_1), \\
\end{align*}
\begin{align*}
\cR_2 \overset{b} \geq \lim_{\eps \to 0}\limsup_{n \to \infty} \frac{ \ell_{\denc}^{\eps}(\bY_1^n)}{n} \geq H(Y_1|X_1), \\
\end{align*}
\begin{align*}
\cR_1+\cR_2 &\overset{c}\geq \lim_{\eps \to 0}\limsup_{n \to \infty} \left[ \frac{ \ell_{\denc}^{\eps}(\bX_1^n) + \ell_{\denc}^{\eps}(\bY_1^n)}{n} \right]\\
& \hspace{1mm} \geq H(X_1,Y_1),
\end{align*}
 where $a$, $b$ follow from the Definition \ref{rate} and $c$ follows because
\begin{align*}
&\lim_{\eps\to 0}\limsup_{n \to \infty} \frac{ \ell_{\denc}^{\eps}(\bX_1^n)}{n}+\lim_{\eps\to 0}\limsup_{n \to \infty} \frac{ \ell_{\denc}^{\eps}(\bY_1^n)}{n} \geq\\ 
& \lim_{\eps\to 0}\limsup_{n \to \infty} \left[ \frac{ \ell_{\denc}^{\eps}(\bX_1^n) + \ell_{\denc}^{\eps}(\bY_1^n)}{n} \right].
\end{align*}

\section{Multiple Access Channel}

\begin{definition}
A two user discrete multiple-access channel consists of three alphabets  $\cX$, $\cY$, $\cZ$ and a probability transition matrix
$P_{Z|X,Y}$.
\end{definition}

\begin{definition}
A $(2^{C_{1}^{\eps}(\cM^{(1)})}, 2^{C_{2}^{\eps}(\cM^{(1)})})$ one-shot code for the two user multiple-access channel consists of two message
sets $\cM^{(1)}= [1:2^{C_{1}^{\eps}(\cM^{(1)})}]$, $\cM^{(2)} = [1:2^{C_{2}^{\eps}(\cM^{(2)})}]$ and a random variable $U$ with range $\cU$, two encoding functions, 
\begin{align}
\label{enc1}
e_{\cM^{(1)}} : \cM^{(1)} & \rightarrow \cX,\\
\label{enc2}
e_{\cM^{(2)}} : \cM^{(2)} & \rightarrow \cY,
\end{align}
and a decoding function 
\beq
g : \cZ   \to \cM^{(1)} \times \cM^{(2)}. \nonumber
\enq
We will call 
\begin{equation*}
 C_{1}^{\eps}(\cM^{(1)}) = \log|\cM^{(1)}|,
 \hspace{3mm}C_{2}^{\eps}(\cM^{(1)} )= \log|\cM^{(2)}|
\end{equation*}
 the coding rates of the encoders $1$ and $2$.
\end{definition}

\begin{definition}
An $\eps$ one-shot rate pair, denoted by $(\cR_{1}^{\eps}, \cR_{2}^{\eps})$, is said to be achievable for the two user multiple-access channel if and only if there exists a code $(2^{C_{1}^{\eps}(\cM^{(1)})}, 2^{C_{2}^{\eps}(\cM^{(1)})})$ such that 
\begin{align}
\label{errb}
\Pr\{g(Z) &\neq (M_1, M_2)\} \leq \eps,\\
\cR_{1}^{\eps} & \leq C_{1}^{\eps}(\cM^{(1)}),\\
\cR_{2}^{\eps} & \leq C_{1}^{\eps}(\cM^{(2)}).
\end{align}
\end{definition}

\begin{definition}
A $(2^{C_{1}^{\eps}(\cM^{(1)}_n)}, 2^{C_{2}^{\eps}(\cM^{(1)}_n)})$ $n$-shot code for the two user multiple-access channel consists of two message
sets $\cM^{(1)}_n= [1:2^{C_{1}^{\eps}(\cM^{(1)}_n)}]$, $\cM^{(2)}_n = [1:2^{C_{2}^{\eps}(\cM^{(2)}_n)}]$ and a random variable $U$ with range $\cU$, two encoding functions, 
\begin{align*}
e_{\cM^{(1)}_n} : \cM^{(1)}_n  & \rightarrow \cX^n,\\
\label{enc2}
e_{\cM^{(2)}_n} : \cM^{(2)}_n  & \rightarrow \cY^n,
\end{align*}
and a decoding function 
\beq
g : \cZ^n  \to \cM^{(1)}_n \times \cM^{(2)}_n. \nonumber
\enq
where $\cX^n, \cY^n$ and $\cZ^n$ denote the n-fold Cartesian product of $\cX, \cY$ and $\cZ$. We will call 
\begin{align*}
\frac{C_{1}^{\eps}(\cM^{(1)}_n)}{n} &= \frac{\log|\cM^{(1)}_n|}{n},\\
\frac{C_{2}^{\eps}(\cM^{(2)}_n)}{n} &= \frac{\log|\cM^{(2)}_n|}{n},\\
\end{align*}
as the coding rate for encoders $1$ and $2$.
\end{definition}

\begin{definition}
\label{memoryless}
Let $x^n \in \cX^n$ and $y^n \in \cY^n$ be two $n$ length input sequences of a two user multiple-access channel and let $z^n\in \cZ^n$ be the output sequence of the channel. 
We call the channel to be memoryless if 
\beq
P_{Z^n|X^n,Y^n}(z^n|x^n,y^n) = \Pi_{i=1}^{n}P_{Z|X,Y}(z_i|x_i,y_i).
\enq
\end{definition}

We now give the definition of rate pair which is achievable asymptotically for the two user multiple access channel.
\begin{definition}
\label{rate}
Rate pair $(\cR_1, \cR_2)$ is achievable $\iff$ There exists a triplet $(e_{\cM^{(1)}_n}, e_{\cM^{(2)}_n}, g)$, such that

\begin{align*}
\Pr\left\{g(Z^n,U)\}\neq (M_1, M_2)\right\}
& \leq \eps, \\
\lim_{\eps \to 0}\liminf_{n \to \infty}\frac{C_{1}^{\eps}(\cM^{(1)}_n)}{n} & \geq \cR_1, \\
\lim_{\eps \to 0}\liminf_{n \to \infty}\frac{C_{2}^{\eps}(\cM^{(2)}_n)}{n} & \geq \cR_2.
\end{align*}

\end{definition}

\begin{definition}
\label {deltaconst} 
For $(X,Y,Z) \sim P_{XYZ}$, we define the following sets
where

\begin{align*}
\cA_{\delta}(P_{XYZ}) &:= \{(x,y,z) : \mximin{XYZ} \leq -\log P_{XYZ}(x,y,z)  \\& \hspace{6mm}\leq \mximax{XYZ}, x\in \cA_{\delta}(P_{XY}) \cap  \cA_{\delta}(P_{XZ}) ,\\
&\hspace{6mm} z  \in \cA_{\delta}(P_{XZ}) \cap  \cA_{\delta}(P_{YZ}),  y \in \cA_{\delta}(P_{XY}) \cap \\
& \hspace{6mm}  \cA_{\delta}(P_{YZ}) \},                       
\end{align*}
where
\begin{align*}
\mximin{XYZ} & := H(XYZ) - |H(XYZ) - H_{\infty}^{\delta}(XYZ)| \\
& \hspace{6mm} - \delta (\log|\cX|+\log|\cY|+\log|\cZ|), \\
\mximax{XYZ} & := H(XYZ) + | H(XYZ) - H_{-\infty}^{\delta}(XYZ) | \\
& \hspace{6mm}+ \delta (\log|\cX|+\log|\cY|+\log|\cZ|),
\end{align*}
and $\cA_{\delta}(P_{XY})$, $\cA_{\delta}(P_{XZ})$ and $\cA_{\delta}(P_{YZ})$ are defined in similar way as \eqref{const. XY}.
\end{definition}

We give one-shot bounds for the transmission rates of the user 1 and user 2 for the
multiple-access channel in the following theorem.

\begin{theorem}
\label{Macbounds}
Let $(X,Y,Z) \sim P_{XYZ}$ and $\eps$, $\eps_i \in \mathbb{R}^{+}$, $i=0,1,2,3$ with $\sum_{i=0}^3 \eps_i \leq \eps$. \\


\noindent {\bf Achievability}: There exists a one-shot communication protocol for the
multiple-access channel for a rate pair $(C_1^{\eps}, C_2^{\eps})$
satisfying the following conditions

\begin{align}
\label{ax_1}
C_1^{\eps} &\leq \mximin{X}+\mximin{YZ}- \mximax{XYZ} \nonumber\\ 
& \hspace{6mm}+\log(\eps_1), \\
\label{ax_2}
C_2^{\eps} &\leq \mximin{Y}+\mximin{XZ}- \mximax{XYZ} \nonumber \\
& \hspace{6mm}+\log(\eps_2), \\
\label{ax_1,ax_2}
C_1^{\eps}+C_2^{\eps} &\leq \mximin{X}+\mximin{Y}+\mximin{Z}- \nonumber \\
&\hspace{6mm} \mximax{XYZ}+\log(\eps_3).
\end{align}
where $\delta$ is chosen such that
\beq
\Pr \left\{ (X, Y, Z) \notin \cA_\delta(P_{XYZ}) \right\} \leq \eps_0.
\enq
\end{theorem}

%
%
%
 
 \begin{proof}

We show the existence of a one-shot communication protocol for the
multiple-access channel (based on two-universal hash functions) and is given by the 
following steps. Fix a $P_{XY} = P_XP_Y$.    

 \begin{enumerate}
 
 \item{\bf{Codebook generation}}. Randomly and independently generate $ 2^{C_1^{\eps}}$ codewords $x(m_1)$, $m_1 \in[1:2^{C_1^{\eps}}]$, each according to $P_X$. Similarly generate
 $2^{C_2^{\eps}}$ codewords $x(m_2)$, $m_2 \in[1:2^{C_2^{\eps}}]$, each according to $P_Y$. These codewords form the codebook, which is revealed to the senders and the receiver.
 
 \item{\bf{Encoding}}. To send a message $m_1$, sender $1$ sends the codeword $X(m_1)$. Similarly, to send $m_2$ sender $2$ sends $Y(m_2)$. 
 
 \item {\bf{Decoding}}. The receiver $Z$ chooses the pair $(m_1,m_2)$ such that 
 \beq
 (x(m_1), y(m_2), z) \in  \cA_{\delta}(P_{XYZ})
 \enq
 if such a pair exists and is unique; otherwise, an error is declared.
 
 \item{\bf{Probability of error analysis}}. By the symmetry of the random code construction, the conditional probability of error
 does not depend on which pair of messages is sent. Thus, without loss of generality, we assume that the message pair $(m_1,m_2)$ was sent. \\
 
Let $\eps_0, \eps_1, \eps_2, \eps_3 \in \mathbb{R}^+$ and $\sum_{k=0}^3 \eps_k \leq \eps$.
The probability of error of the above protocol is computed by defining the following events. 

\begin{align*}
E_0 & := \{(X(m_1),Y(m_2),Z) \notin \cA_{\delta}(P_{XYZ}) \},\\
E_{ \acute{m_1}, \acute{m_2}} & := \{\exists( ~ \acute{m_1}, \acute{m_2}) \neq (m_1,m_2) : (X( \acute{m_1}),\\
& \hspace{6mm}Y( \acute{m_2}),Z)  
\in \cA_{\delta}(P_{XYZ}) \}.
\end{align*}
 \end{enumerate}

The probability of error is given by the probability of the union of these events and we upper bound that by the
union bound as
\begin{align*}
P_e & = \Pr \left\{ E_0 \bigcup \cup_{( ~ \acute{m_1}, \acute{m_2}) \neq (m_1,m_2)} E_{ ~ \acute{m_1}, \acute{m_2}} \right\} \\
         & \leq \Pr\{ E_0 \}  ~~~+  \sum_{ ~ \acute{m_1}\neq m_1, ~ \acute{m_2}=m_2} \Pr\{ E_{ ~ \acute{m_1},m_2}\} \\
     &\hspace{6mm}+ \sum_{ ~ \acute{m_1} = m_1, ~ \acute{m_2} \neq m_2} \Pr\{ E_{m_1,~ \acute{m_2}}\}\\
&\hspace{6mm} +\sum_{ ~ \acute{m_1}\neq m_1, ~ \acute{m_2} \neq m_2} \Pr\{ E_{ ~ \acute{m_1}, ~ \acute{m_2}}\}.
 \end{align*}
 
For any $\eps_{0}\geq 0$, we could choose $\delta \geq 0$ such that $ \Pr\{ E_0 \} \leq \eps_{0} $.   

For  $~ \acute{m_1}\neq m_1$, we have
\begin{align*}
 \Pr\{ E_{ ~ \acute{m_1},m_2}\} & = \Pr\left\{\left(X(~ \acute{m_1}),Y(m_2),Z\right) \in \cA_{\delta}(P_{XYZ})\right\}\\
                                        & = \sum_{(x,y,z) \in \cA_{\delta}(P_{XYZ})} \Pr(x)\Pr(y,z)\\
                                        & \overset{a}{\leq} | \cA_{\delta}(P_{XYZ})| 2^{ - (\mximin{X}+\mximin{YZ})}\\
                                        & \overset{b}{\leq} 2^{-\left( \mximin{X}+\mximin{YZ}- \mximax{XYZ}\right)},
\end{align*}
 where $a$ and $b$ follows from the properties of $ \cA_{\delta}(P_{XYZ})$.
Similarly for $ ~ \acute{m_2} \neq m_2$, we have 
\beq
 \Pr\{ E_{m_1, ~ \acute{m_2}}\} \leq  2^{-\left( \mximin{Y}+\mximin{XZ}- \mximax{XYZ}\right)} \nonumber,
\enq
                                           
and for $~ \acute{m_1} \neq m_1, ~ \acute{m_2} \neq m_2$,
\beq
 \Pr\{ E_{~ \acute{m_1},~ \acute{m_2}}\} \leq 2^{-\left(\mximin{X}+\mximin{Y}+\mximin{Z}- \mximax{XYZ}\right)}. \nonumber
\enq

Thus,
\begin{align*}
P_e & \leq  \eps_{0} + 2^{C_1^{\eps}}  2^{-\left( \mximin{X}+\mximin{YZ}- \mximax{XYZ}\right)}\nonumber\\ 
     &\hspace{5mm} + 2^{C_2^{\eps}}  2^{-\left( \mximin{Y}+\mximin{XZ}- \mximax{XYZ}\right)} +  2^{C_1^{\eps}+C_2^{\eps}}\nonumber\\
     & \hspace{5mm}  2^{-\left(\mximin{X}+\mximin{Y}+\mximin{Z}- \mximax{XYZ}\right)}.
\end{align*}

Hence, $ \sum_{ ~ \acute{m_1}\neq m_1, ~ \acute{m_2}=m_2} \Pr\{ E_{ ~ \acute{m_1},m_2}\} \leq \eps_{1}$ 
if $C_1^{\eps} \leq \mximin{X}+\mximin{YZ}- \mximax{XYZ}+\log(\eps_1)$.
Similarly, we could show that  $ \sum_{ ~ \acute{m_1}= m_1, ~ \acute{m_2}\neq m_2} \Pr\{ E_{ m_1,~ \acute{m_2}}\} \leq \eps_{2}$
if $C_2^{\eps} \leq \mximin{Y}+\mximin{XZ}- \mximax{XYZ}+\log(\eps_2)$ and
$\sum_{ ~ \acute{m_1}\neq m_1, ~ \acute{m_2}\neq m_2} \Pr\{ E_{ ~ \acute{m_1},~ \acute{m_2}}\}
\leq \eps_{3}$ if $C_1^{\eps}+C_2^{\eps} \leq \mximin{X}+\mximin{Y}+\mximin{Z}- \mximax{XYZ}+\log(\eps_3)$.

The above analysis shows that the average probability of error, averaged over all choices of codebooks in the random code construction achieves the error bound in  \eqref{errb} . Thus, there exists at least one code which achieves the desired error bound in \eqref{errb}. \\
\end{proof}

The asymptotic optimality of the above achievable rate pair for the case of memoryless multiple-access channel can be proved using ideas similar to that used in the proof of Lemma \ref{asymp-sw} and by invoking Definition \ref{memoryless} and Definition \ref{rate}.

\section{Conclusions and Acknowledgement}

In conclusion, we defined a novel one-shot typical set based on smooth entropies which helps us to get the one-shot rate region for distributed source coding and multiple access channel
while leveraging the results from the asymptotic analysis. We parametrize this one-shot typical set using $\delta$. The choice of $\delta$, further depends upon the amount of error ($\eps$) that is allowed. In short, the decoding region is a function of $\delta$. Primarily, the motivations for the definition of this set is inspired
by the definition of typical set for the asymptotic case and it turns out to be a serendipitous
scenario where such a set has the desired properties asymptotically when large
i.i.d. copies are available.

The authors wish to thank R. Renner for giving his helpful comments on the subject.

\bibliographystyle{ieeetr}
\bibliography{master}

\end{document}